# Nearly Lattice Matched GaAs/Pb$_{1-x}$Sn$_x$Te Core-Shell Nanowires

*Sania Dad[1], Piotr Dziawa[1], Wiktoria Zajkowska[1], Sławomir Kret[1], Mirosław Kozłowski[1], Maciej Wójcik[1], and Janusz Sadowski[1,2]*

[1]Institute of Physics, Polish Academy of Sciences, Aleja Lotnikow 32/46, PL-02-668 Warsaw, Poland

[2]Department of Physics and Electrical Engineering, Linnaeus University, SE-391 82 Kalmar, Sweden

## ABSTRACT

We investigate the full and half-shells of Pb$_{1-x}$Sn$_x$Te topological crystalline insulator deposited by molecular beam epitaxy on the sidewalls of wurtzite GaAs nanowires (NWs). Due to the distinct orientation of the IV-VI shell with respect to the III-V core the lattice mismatch along the nanowire axis is less than 4%. The Pb$_{1-x}$Sn$_x$Te solid solution is chosen due to the topological crystalline insulator properties for some critical concentrations of Sn (x ≥ 0.4). The IV-VI shells are grown with different compositions spanning from binary SnTe, through Pb$_{1-x}$Sn$_x$Te with decreasing $x$ value down to binary PbTe (x = 0). The samples are analyzed by scanning transmission electron microscopy, which reveals the presence of (110) or (100) oriented binary PbTe and (100)Pb$_{1-x}$Sn$_x$Te on the sidewalls of wurtzite GaAs NWs.

## 1. Introduction:

The combination of semiconductor heteroepitaxial systems without generation of numerous defects often remains inaccessible due to large lattice mismatch between materials of interest [1-3]. However, in materials stacked in the nanostructures the lattice matching requirements indispensable in the case of large area planar heterostructures are much less severe. In the nanoscale dimensions, the stress/strain can be shared between both components of the heterostructures, in contrast to the case of thin layers grown on substrates usually thicker by several orders of magnitude [4, 5]. In this report, we present a successful attempt of combining in the core-shell nanowire heterostructures, two much dissimilar materials (with the lattice misfit $f$ = 12 to 14%) namely well-known III-V semiconductor GaAs in the wurtzite (WZ) crystalline phase (occurring in the nanowires only) and a narrow bandgap IV-VI semiconductor (Pb$_{1-x}$Sn$_x$Te solid solution) exhibiting also topological crystalline insulator (TCI) properties.

In the last decade Pb$_{1-x}$Sn$_x$Te was identified to be a topological crystalline insulator (TCI). A topological insulator is a quantum material that has protected gapless electronic states on its surface or edges but the bulk electronic states are gaped like in a conventional insulator [6]. TCIs metallic surface states are protected by crystal symmetry in contrast to the time-reversal symmetry protection of formally known Z$_2$ topological insulators (TIs) [6, 7]. Apart from topological properties, narrow bandgap IV-VI semiconductors possess other unique features making them promising materials for applications in thermoelectric and mid-infrared optoelectronic devices such as photodetectors, thermoelectric coolers, infrared light sources (lasers), and detectors [8, 9].

Investigation of the IV-VI narrow bandgap lead-tin chalcogenides has gained momentum over the last decade due to discovery of topologically protected surface states(TSS) on high symmetry surfaces. The indispensable condition for the occurrence of TSS, namely mirror symmetry of crystalline lattice and electronic band inversion are fulfilled in certain binary IV-VI alloys [10]. Shortly after the first theoretical report on the TCI phase [11], the topological phase of SnTe protected by the crystalline symmetry has been theoretically predicted [12] and confirmed experimentally for SnTe [13], $Pb_{1-x}Sn_xTe$ [14] and $Pb_{1-x}Sn_xSe$ [15] solid solution. To maximize the effects related to the topologically protected states [16, 17] quasi one-dimensional nanowires (NWs) are beneficial due to their high surface to volume ratio. Moreover. NWs enable the exploration of novel phenomena such as topologically protected 1D hinge states emerging along the 3 edges [18]. Several surfaces hosting topologically protected states in individual NWs – four surfaces hosting topologically protected states at the NW sidewall surfaces of native IV-VI NWs (with square cross-sections [18]) and six such surfaces for GaAs-IV-VI core-shell NWs (with hexagonal cross-sections [19]) described here, can occur in the NWs.

Over two decades of research on semiconducting NWs paved the ways for highly controllable composition, length, diameter, ordered positioning and crystal structure of NWs [20-25]. One of the basic methods for the NWs growth exploits the vapor-liquid-solid (VLS) mechanism which was identified by Wagner and Ellis almost a half-century ago [26]. This technique is still an excellent approach to obtain high-quality NWs [25, 27]. A typical VLS mechanism requires a liquid nano-droplet (catalyst), usually keeping the size of the droplet as required for the diameter of NWs, which acts as a nucleation center for the growth. In a VLS technique the size and wetting angle of liquid nano-droplet determines the diameter and crystalline structure of a NW [28]. This approach has been successfully applied to obtain NWs of II-VI, III-V, and IV-VI semiconductors [29-31].

In comparison to simple NWs, heterostructures offer quite some new functionalities [32-37], especially core-shell NWs offer a wide range of applications due to their numerous advantages for altering the material properties. For example, the band gap of the core can be tuned by straining the nanowire using lattice mismatched shell [38]. Auger recombination and surface trap states are reduced in core-shell structures, the appropriately designed shells can also improve optical and charge transport properties of the core material [32]. Carrier multiplication [39] phenomenon can lead to promising performances in photodetectors, solar cells, and light-emitting diodes [40, 41]. Recently, the PbTe/PbS 1D core/shell nanostructures were reported in which transport of electrons was prevailing over the ambipolar one. Such a property makes them a suitable candidate for thermoelectricity and photovoltaic devices. Moreover, core-shell NWs of appropriate size and composition were also investigated as electro-catalysts for oxygen reduction reaction [42-45].

The IV-VI materials are grown usually on the native rock-salt or barium fluoride substrates which, however, results in a poor mechanical stability, unfavorable thermal performance, and difficulties in processing due to the reactivity of the substrate with water. Therefore, the growth of IV-VI (rock-salt) on Si and III-V or II-VI (zinc-blende) substrates is preferred. This approach offers several merits such as better surface morphology and crystalline perfection, chemical and mechanical stability, and thermal conductivity as compared to native rock-salt or fluorite substrates [46-48]. Moreover, in the case of NWs even the substantial lattice mismatch of IV-VI materials to Si or GaAs is not an obstacle for the growth due to their small lateral dimensions. The reduced dislocation densities at interfaces can be obtained in heterostructures with lattice mismatched core-shell NWs [49]. Inspired by these facts, the IV-VI ($Pb_{1-x}Sn_xTe$) both shells and half-shells are deposited on the sidewalls of III-V GaAs NW cores to obtain (in further research perspective) the expected higher-order topological insulator (HOTI) states at the NW edges [18]. In such a hybrid configuration it is possible to obtain relatively long NWs (contrary to typical MBE-grown IV-VI NWs [27, 50]) which is beneficial for studying their characteristics involving topologically protected electronic states [51].

The core-shell NWs are examined by scanning and transmission electron microscopy (SEM) and (TEM), respectively. The lattice mismatch of the shell with respect to the core is measured on the basis of TEM images and is compared to the theoretical calculations. To our best knowledge, $Pb_{1-x}Sn_xTe$ shells deposited on GaAs NW cores have not been reported before.

## 2. Experimental Section:

### 2.1 Substrate Preparation:

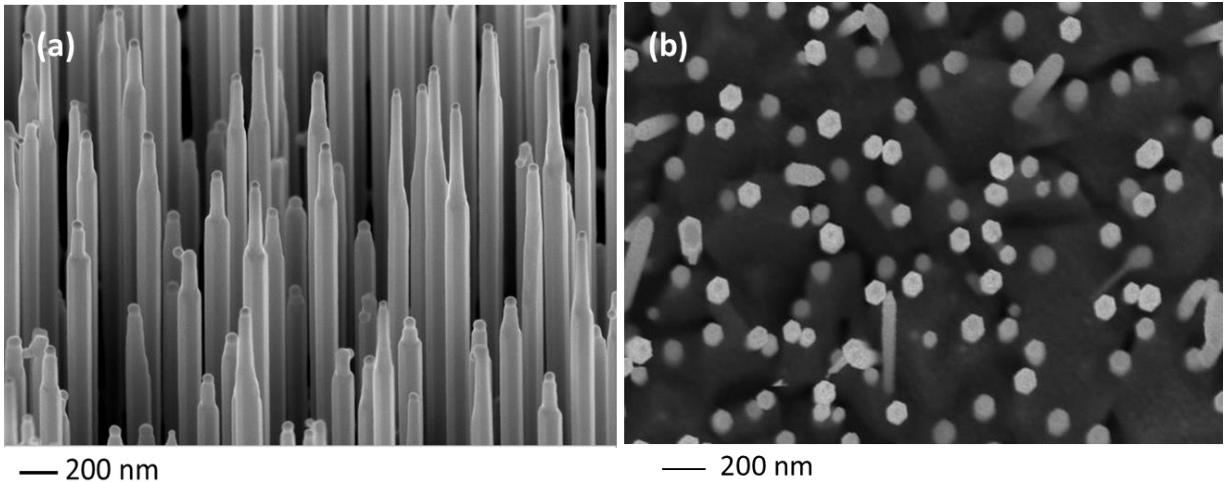

**Figure 1.** The side (a) and top (b) view of WZ GaAs NWs grown by gold-assisted VLS method.

In the first step we have grown of Au-catalyzed GaAs NWs on GaAs (111)B substrates in the III-V MBE system. These NWs have a predominant wurtzite (WZ) crystal structure and grow perpendicularly to the substrate surface, along the [0001] direction (see SEM images shown in Fig. 1). Then, III-V NWs were transferred in air to the separate IV-VI MBE system and used as templates (cores) for IV-VI shells deposition.

## 2.2 Growth Process:

Prior to the deposition of shells, the GaAs NWs were annealed at about 590 °C to thermally desorb native oxides from the sidewalls. In-situ reflection high-energy electron diffraction (RHEED) patterns (see Fig. 2) correspond to mutually oriented NWs perpendicular to the substrate.

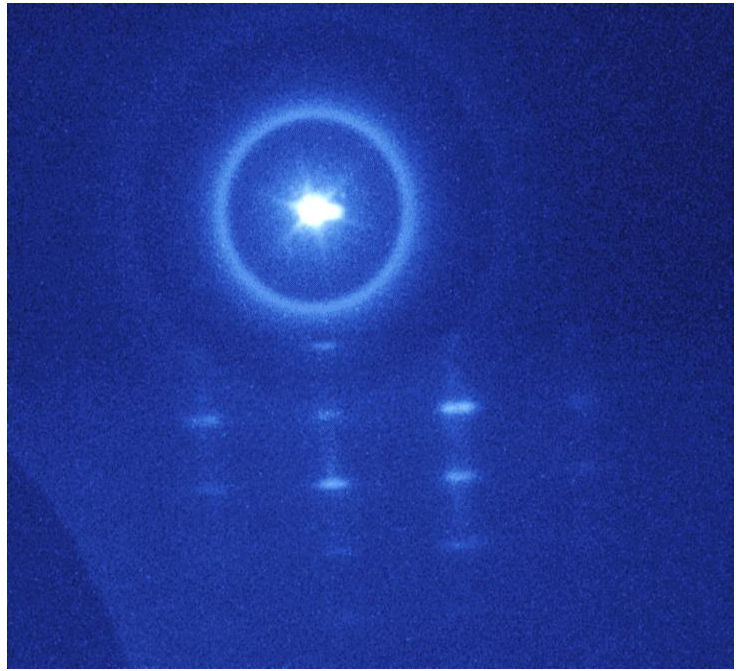

**Figure 2.** RHEED patterns of GaAs NWs after thermal desorption of native oxides in the IV-VI MBE growth chamer.

After annealing, the substrate temperature was decreased down to ($400 < T_{gr} < 450$ °C) to start the growth of the IV-VI shells. In contrast to full-shells, the half-shells are grown in the static conditions, i.e. without substrate rotation. Such a procedure enables the deposition of material on half of the NW, namely on three out of six sidewalls only. $Pb_{1-x}Sn_xTe$ with various chemical compositions of are grown with $x$ ranging from 0 to 1.

### 3. $Pb_1Sn_{1-x}Te$ Shells on Wurtzite GaAs NW:

### 3.1 Binary SnTe shells:

The SEM and TEM images of WZ GaAs NWs with SnTe shells are presented in Fig. 3. A closer look at the entire NWs (Fig 3b) reveals that the shells are non-continuous but relatively locally smooth. The SnTe shells are very well lattice matched to the WZ GaAs NW core in the axial direction due to the specific orientation of both crystalline lattices as explained below. Figure 3c shows the high magnification TEM image of the WZ GaAs/SnTe interface revealing the alignment of the SnTe shell and WZ GaAs NW core crystal lattices. Corresponding electron diffraction images indexing the directions of atomic arrangements are depicted in Fig. 3d. The image of the GaAs/SnTe interface displayed in Fig. 3c is taken in the GaAs [10-10] zone axis. The lattice mismatch between the substrate and the layer described by the misfit parameter $f$ can be calculated using the standard expression:

$$f = (d_{sub} - d_{lay})/d_{lay} \qquad (1)$$

where $d_{sub}$ and $d_{lay}$ are the inter-planar spacings in the substrate and the layer material, respectively. The values of the inter-planar spacings along the specific directions obtained from the TEM analysis are calculated using the following lattice parameters: $c = 6.5701$ Å and $a =$ 3.9845 Å for WZ GaAs [52], and $a = 6.318$ Å for SnTe [53]. these give the distances $d$ between WZ GaAs and SnTe lattice planes: $d_{(0002)}^{\text{GaAs}} = 3.285$ Å and $d_{(200)}^{\text{SnTe}} = 3.159$ Å.

The mismatch between the spacing of (0002) planes of WZ GaAs and the (002) ones of SnTe amounts to:

$$f_{\text{theor}}^{\text{SnTe/GaAs}} = 0.0398 \qquad (2)$$

This means that lattice matching between WZ GaAs and rock-salt SnTe along the NW axis is quite small (about 4%).

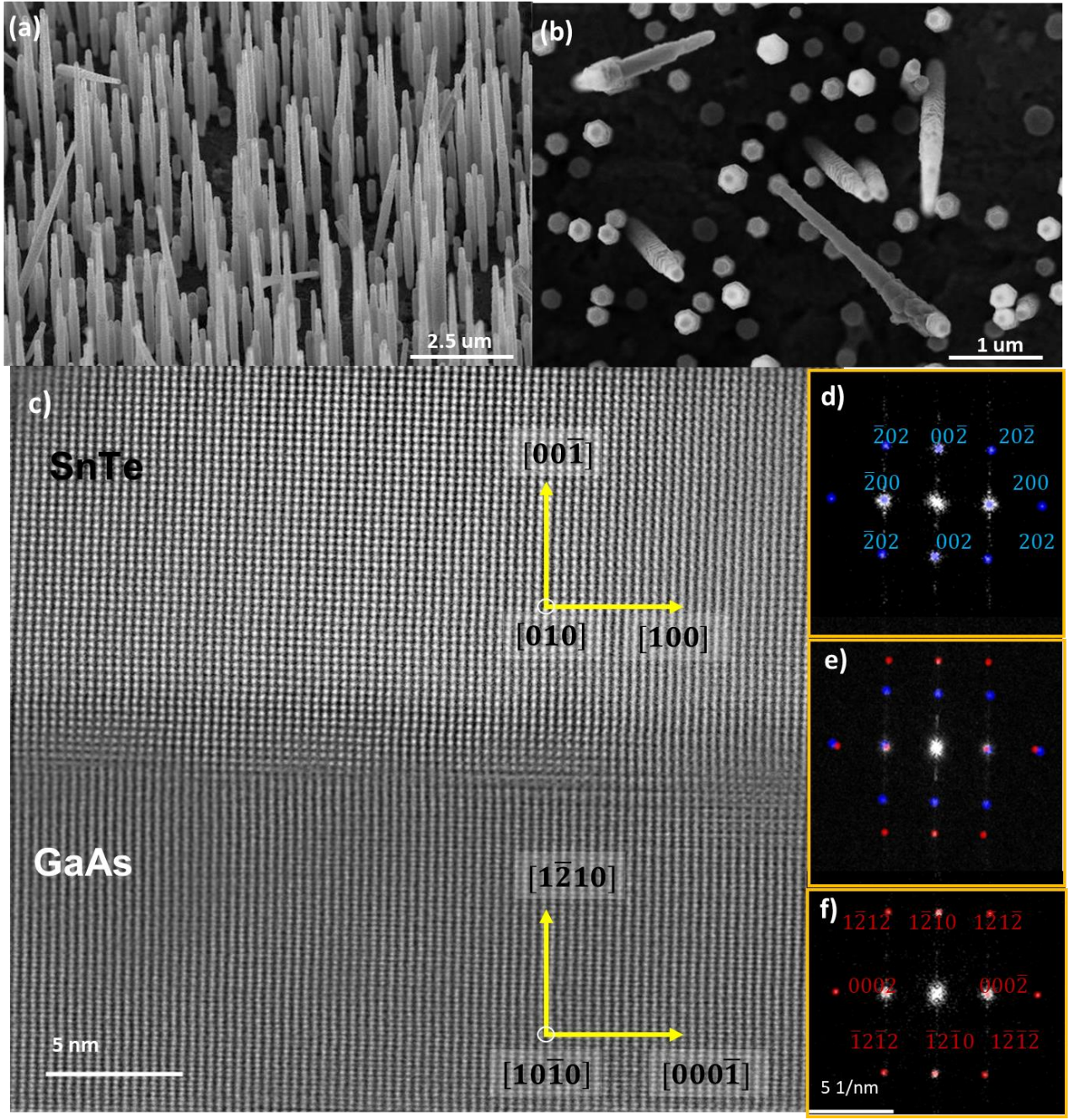

**Figure 3.** (a-b) SEM images (side and top view) of WZ GaAs/SnTe core-shell NW at magnification of 10k and 25k, respectively. (c) STEM image of GaAs(1-210)/SnTe(100) interface along the [010] projection of SnTe with indexed electron diffraction images; shown in the right

panels (d-f) where red dots correspond to GaAs and blue dots correspond to SnTe; the diffraction images share the same scale of 5 nm$^{-1}$.

We can calculate the actual lattice mismatch along the NW axis, $f_{exp}^{Snte/GaAs}$, on the basis of the high-resolution TEM (HR-TEM) images. Considering the fact that the core is much thicker than the shell, we assume the inter-planar spacing value for GaAs to be $d_{(0002)}^{\text{GaAs}} = 3.285$ Å. The inter-planar spacing of SnTe shell measured from the image displayed in Figure 3c is $d_{(200)}^{\text{SnTe}} = 3.174$ Å. With this measured value, the difference in the inter-planar distances between the core and the shell lattices can be calculated using the formula given in equation 1.

$$f_{\text{exp}}^{\text{SnTe/GaAs}} = 0.0349 \qquad (3)$$

This values is 12.3% smaller than theoretical one which indicates that the SnTe shell is partially strained (elongated) along the NW axis with strain value of $\varepsilon_{(002)} = 0.00475$.

### 3.2 Pb$_{0.65}$Sn$_{0.35}$Te shells:

Another type of the IV-VI shells on WZ GaAs NW cores studied here constitute from Pb$_{1-x}$Sn$_x$Te solid solution with Sn content x=0.35. For such a concentration of Sn, the topological phase transition can occur in Pb$_{1-x}$Sn$_x$Te at low temperatures [54]. In the SEM images shown in Fig. 4 it can be seen that the shells are smooth and slightly broken on the inclined NWs contrary to the massively broken shells on the NWs perpendicular to the substrate surface. The high magnification picture of the core-shell structures and their interface is shown in Figure 5a along the [1010] projection of GaAs NW. The well visible fringes in Figure 5a are recognized as a moiré. The right yellows panels in Figure 5(b-d) shows the FFT image of the core, core-shell and the shell respectively. For calculation of the lattice mismatch the lattice parameter of Pb$_{1-x}$Sn$_x$Te for x=0.35 amounts to $a_0$=6.4116 Å at room temperature, as calculated according to the Vegard's law [55]. The inter-planar spacing (d) values for GaAs (0002) and Pb$_{1-x}$Sn$_x$Te (200) are calculated as 3.285 Å and 3.2058 Å respectively. The lattice misfit due to the difference in the lattice parameters of the core and the shell is calculated using the same expression given in Eq (1).

The lattice mismatch along the [1010] projection of GaAs NW axis:

$$f_{\text{theor}}^{\text{PbSnTe/GaAs}} = 0.0247 \qquad (4)$$

The inter-planer spacing of Pb$_{1-x}$Sn$_x$Te shell measured from the high-resolution TEM image is

$$d_{(200)}^{\text{PbSnTe}} = 3.207 \text{ Å}$$

$$f_{\text{exp}}^{\text{PbSnTe/GaAs}} = 0.0243 \qquad (5)$$

Hence the actual misfit is lower and Pb$_{1-x}$Sn$_x$Te is strained along the axial direction by

$$\varepsilon = 0.0245$$

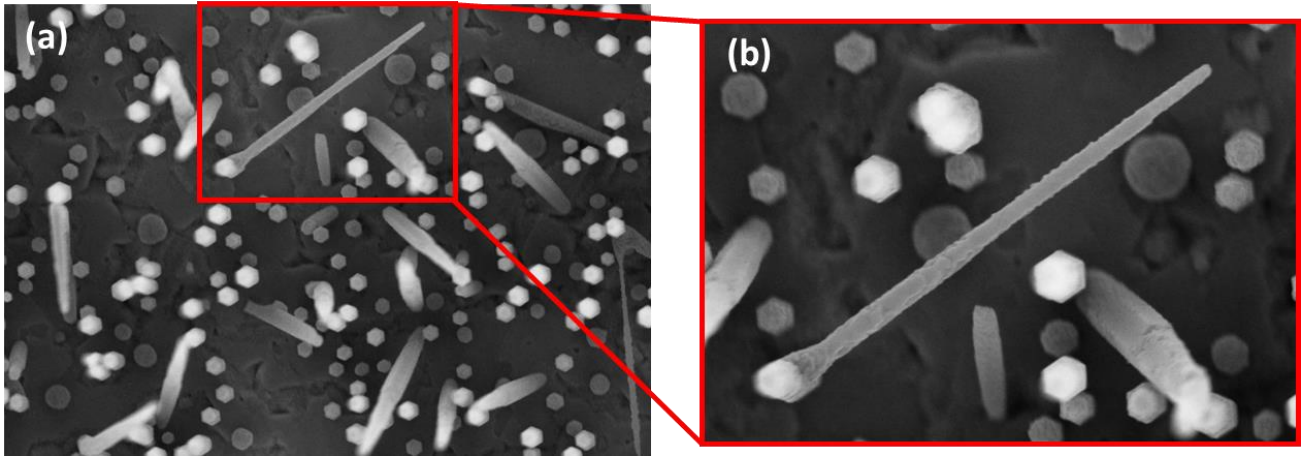

**Figure 4.** (a) SEM images of the WZ GaAs/Pb$_{1-x}$Sn$_x$Te core-shell NWs. Inclined nanowires have smoother shells than the ones perpendicular to the GaAs(111)B substrate surface.

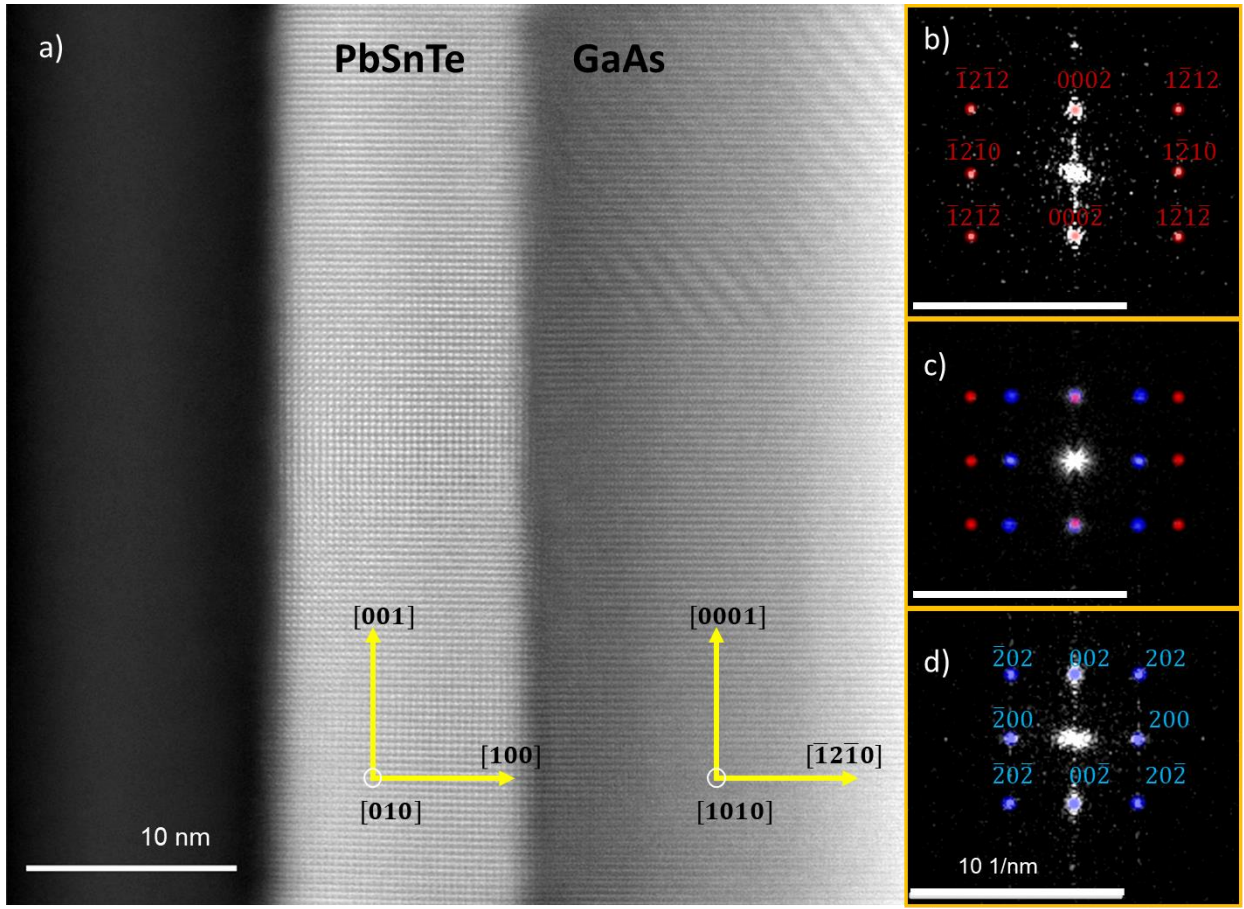

**Figure 5.** (a) STEM image of GaAs(-12-10)/Pb$_{0.65}$Sn$_{0.35}$Te(100) interface along the [010] projection of the shell. The FFT images of the core - (b), interface - (c) and the shell - (d) are presented in the panels with yellow frames, where red dots correspond to the core and blue ones to the shell sharing the same scale of 10 nm$^{-1}$.

### 3.3 Binary PbTe shells:

PbTe shells on GaAs NWs are grown using elemental sources of Pb and Te. Figure 6 shows the SEM images of such core/shell NWs. The SEM images reveal that NWs inclined to the substrate surface (see Fig. 6b) have very smooth and continuous shells in contrast to those on the perpendicular NWs shown in Fig. 6a.

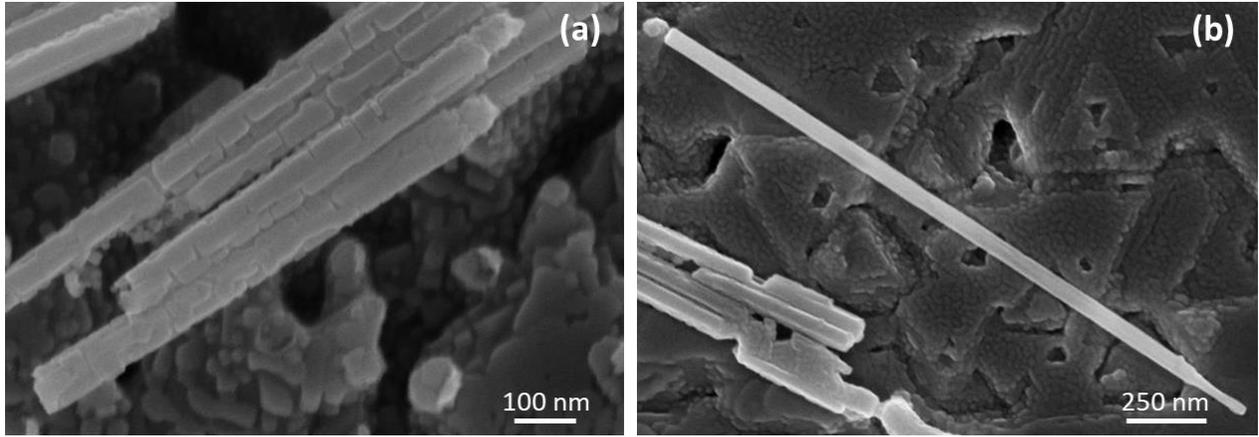

**Figure 6**. SEM images of PbTe shells on (a) - perpendicular and (b) - inclined GaAs NWs. Inclined NW has continuous PbTe shell all along it axis.

For GaAs-PbTe core-shell NWs, w two type of orientations are obtained, depending on the type of WZ GaAs NW facet, as shown in Fig. 7 and Fig. 8. Figure 7 shows TEM images along the [1010] projection of GaAs NW. Figure 7a presents a low magnification image of a broken but locally smooth single core-shell NW. The incoherence in the shell may occur due to the imperfect conditions of the growth such as temperature or mobility of adatoms along the NW sidewalls. Figures 7(c-e) show the indexed electron diffraction images for the NW core indicated by the red dots, the core-shell labelled by red and blue dots and the shell by blue dots. While Fig. 8 shows TEM results of the core-shell NW along the [11-20] projection of the core. Here on the {1-100} WZ GaAs NW side facets the PbTe shell grows in (110) orientation. The growth of (110) PbTe on GaAs has not been reported before, to our knowledge. The schematic representations of the WZ GaAs NW with both types of sidewall facets occurring in WZ phase are shown in Figure 9. It can clearly be seen that the {-1100} WZ GaAs NW sidewalls (*m* WZ planes) are matched to the {110} planes of PbTe, while the {11-20} WZ *a* planes are better matched to the{100} PbTe ones.

The lattice mismatch between the shell and the core is calculated using the same expression given in Eq. (1).

Along the [0001] GaAs NW axis, parallel to the edge of NW sidewall the spacing of (0-10) PbTe planes: $d_{(0-10)}^{PbTe} = 3.23$ Å.

$$f_{theor}^{PbTe/GaAs} = 0.0170 \qquad (6)$$

At such low lattice mismatch, less than 2%, the occurrence of stacking faults is almost negligible. This can be seen in HR-TEM images (Fig 7b). The lattice mismatch calculated from the TEM images is $d_{(0-20)}^{PbTe} = 3.219$ Å

$$f_{exp}^{PbTe/GaAs} = 0.0205 \qquad (7)$$

Along the [11-20] direction of GaAs NW sidewall $d_{(11-20)}^{GaAs} = 1.9922$ Å $d_{(220)}^{PbTe} = 2.2839$ Å.

$$f_{theor}^{PbTe/GaAs} = 0.1365 \qquad (8)$$

And the experimental value of the inter-planar spacing taken from TEM image is $d_{(220)}^{PbTe} = 2.245$ Å

$$f_{exp}^{PbTe/GaAs} = 0.1125 \qquad (9)$$

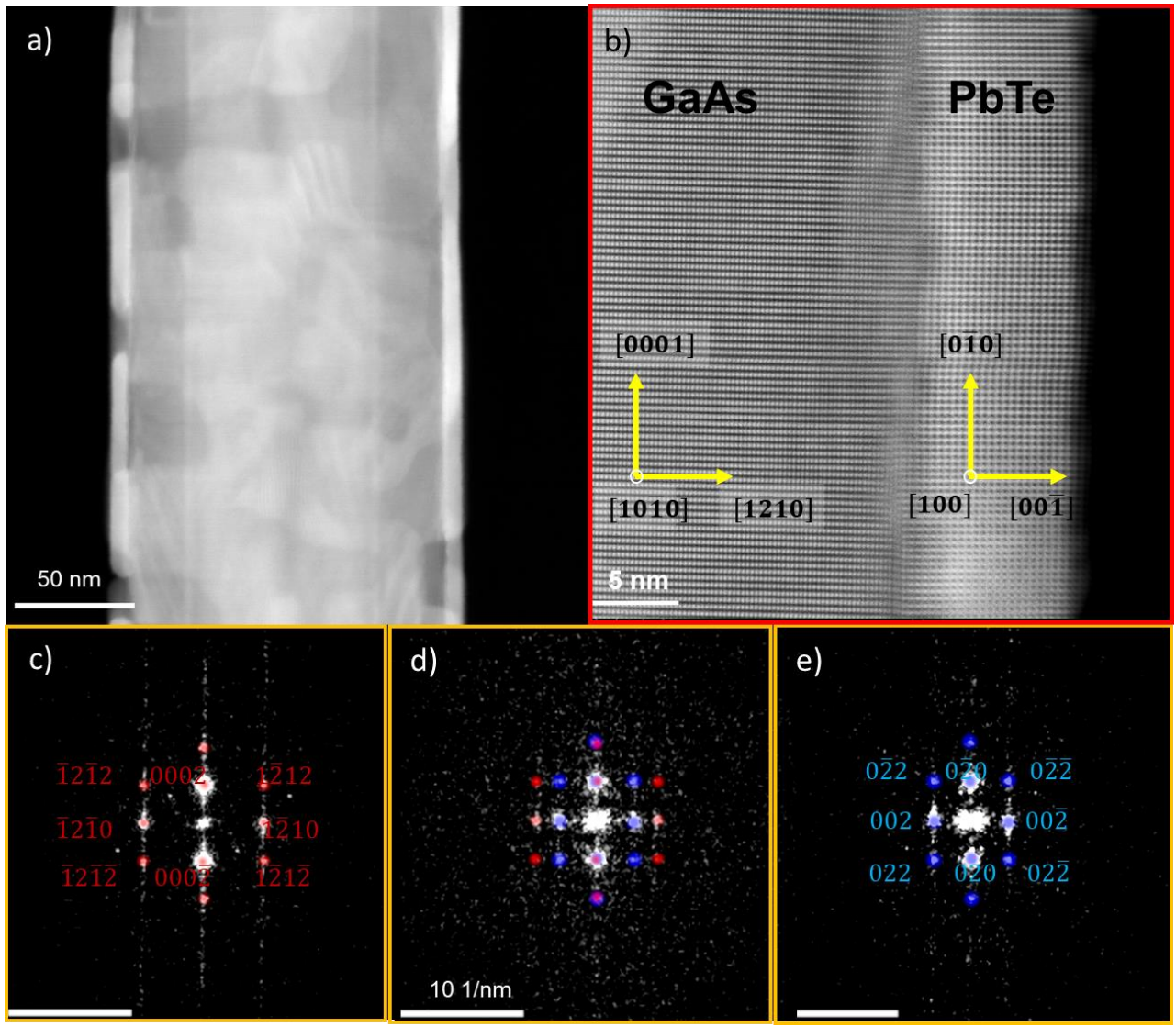

**Figure 7.** TEM images of (001) PbTe shells on WZ GaAs NW with {1-210} sidewalls. (a) -single core-shell NW with locally smooth but non-continuous PbTe shell (b) - a magnified image of a smooth core-shell interface scales along the [100] projection of PbTe. Lower panels show indexed electron diffraction images of (c) - GaAs and (d) - GaAs-PbTe interface and (e) - PbTe.

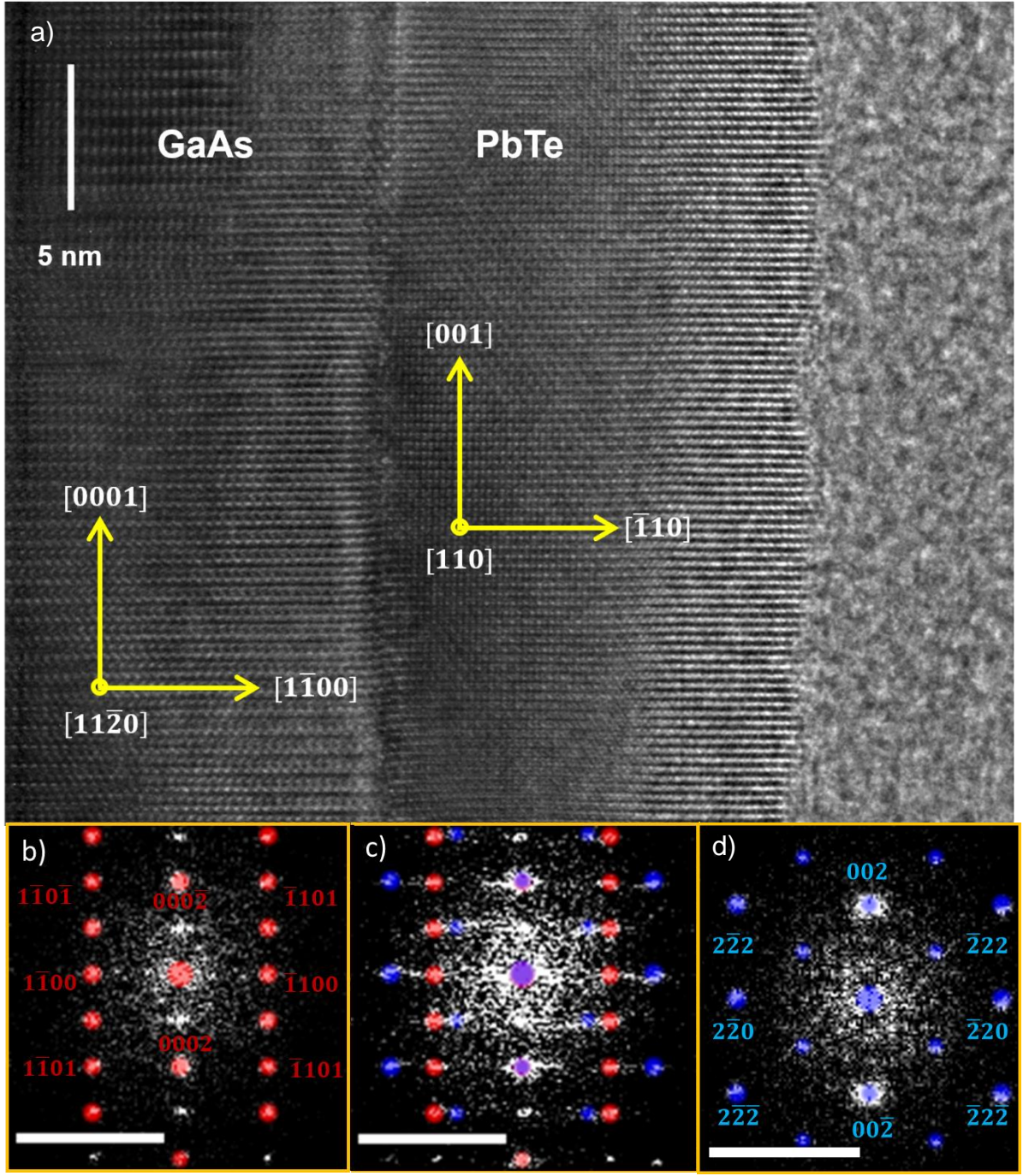

**Figure 8.** TEM images of (110) PbTe shells on WZ GaAs nanowire with {1-100} sidewalls. (a) -
magnified image of the core-shell interface. (c-d) - indexed electron diffraction (FFT) images of
GaAs, GaAs-PbTe and PbTe, respectively.

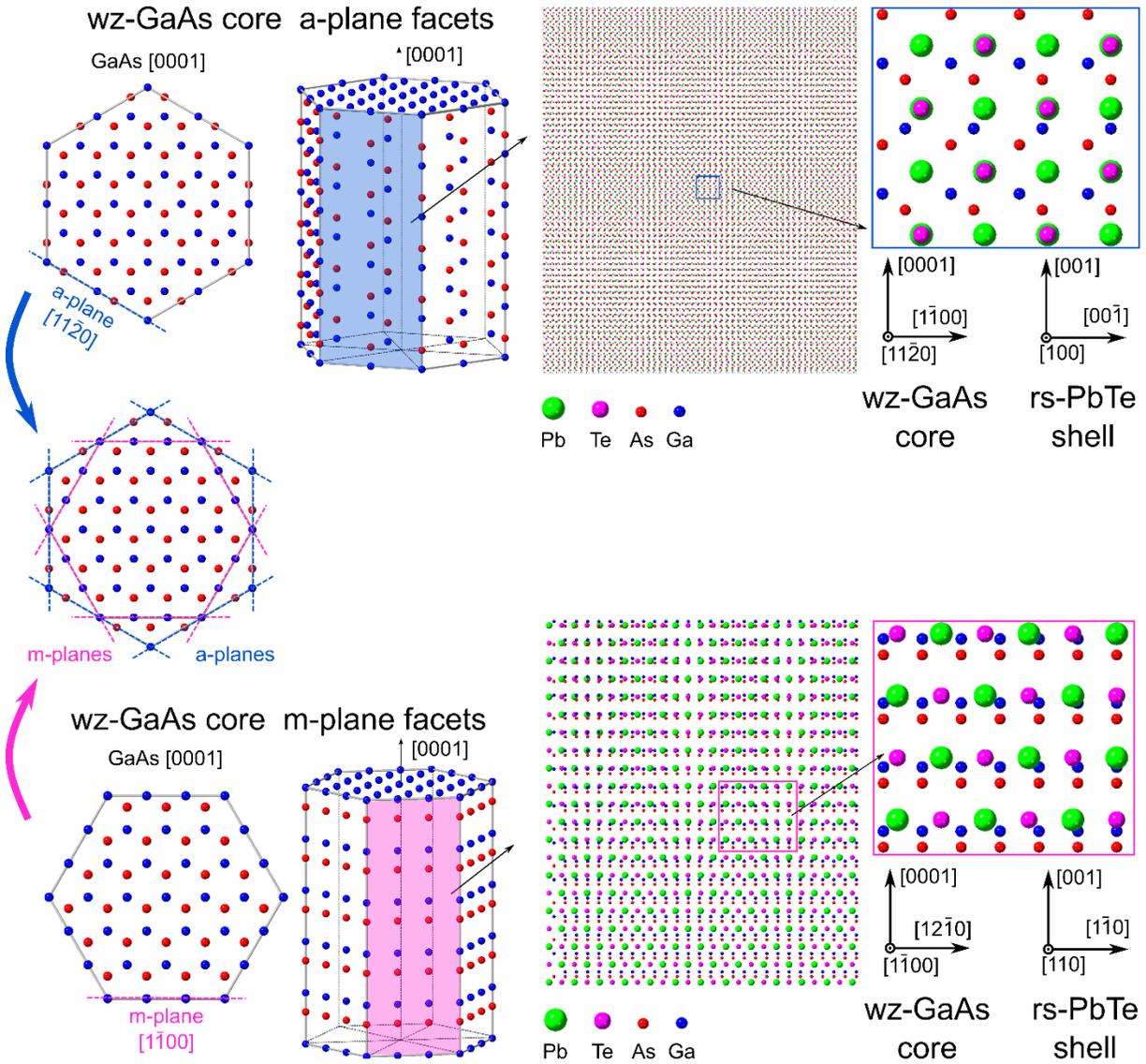

**Figure 9.** Schematic representation of the WZ GaAs nanowire with two types of sidewall facets. Upper panel - {11-20} WZ GaAs sidewalls (*a* planes); lower panel {-1100} sidewalls (WZ *m* planes). The (100) PbTe shells are best fitted to the {11-20} WZ GaAs sidewalls; whereas the (110) PbTe shells are best fitted to the {1-100} ones.

It can be concluded that the PbTe shell is a nearly lattice matched to the GaAs nanowire having a lattice mismatch less than 2%. Moreover, as shown in Fig. 9, the two types of sidewall facets occurring in WZ GaAs NW, namely {-1100} or {11-20} induce the distinct orientation of the IV-VI shells. For PbTe (and $Pb_{1-x}Sn_xTe$) grown on {-1100} WZ GaAs NW facets, the best lattice matching occurs for {110} orientation of the IV-VI shell planes, and for {11-20} WZ GaAs core facets the $Pb_{1-x}Sn_xTe$ {100} shell planes fit best.

### 4. $Pb_1Sn_{1-x}Te$ Half-Shells on WZ GaAs NWs:

The core-(half-shell) of $Pb_{1-x}Sn_xTe$ (for x = 0.56) are grown without substrate rotation. The defect-free $Pb_{1-x}Sn_xTe$ half shells shown in Figure 10 (a) have the same crystalline perfection as the core NW (see Fig.10d).

The theoretical lattice mismatch along the NW axis is calculated using the p interlayer spacing parameters

$d^{GaAs}_{(0002)} = 3.285$ Å and $d^{PbSnTe}_{(020)} = 3.194$ Å

$f^{PbSnTe/GaAs}_{theor} = 0.0284$ \hspace{1cm} (10)

The lattice mismatch in the direction perpendicular to the NW axis but parallel to the sidewall (along the [10-10] projection of the GaAs NW axis, can be evaluated from the cross-sectional high-resolution TEM image of the NW with the $Pb_{1-x}Sn_xTe$ half shell shown in Fig. 10b. The spacing of $Pb_{1-x}Sn_xTe$ (020) planes measured from the NW cross-section image $d^{PbSnTe}_{(020)} = 3.221$ Å

The measured mismatch between $Pb_{1-x}Sn_xTe$ (020) and GaAs (0002) is equal to:

$f^{PbSnTe/GaAs}_{exp} = 0.0198$ \hspace{1cm} (11)

Consequently, the strain of $Pb_{1-x}Sn_xTe$ lattice in the [010] direction:

$\varepsilon_{(020)} = 0.0196$

The successful growth of smooth and unbroken half-shells of Pb$_{1-x}$Sn$_x$Te on GaAs NWs can enable investigation of their expected higher-order topological crystalline insulator properties (topological hinge states) along the NW corners.

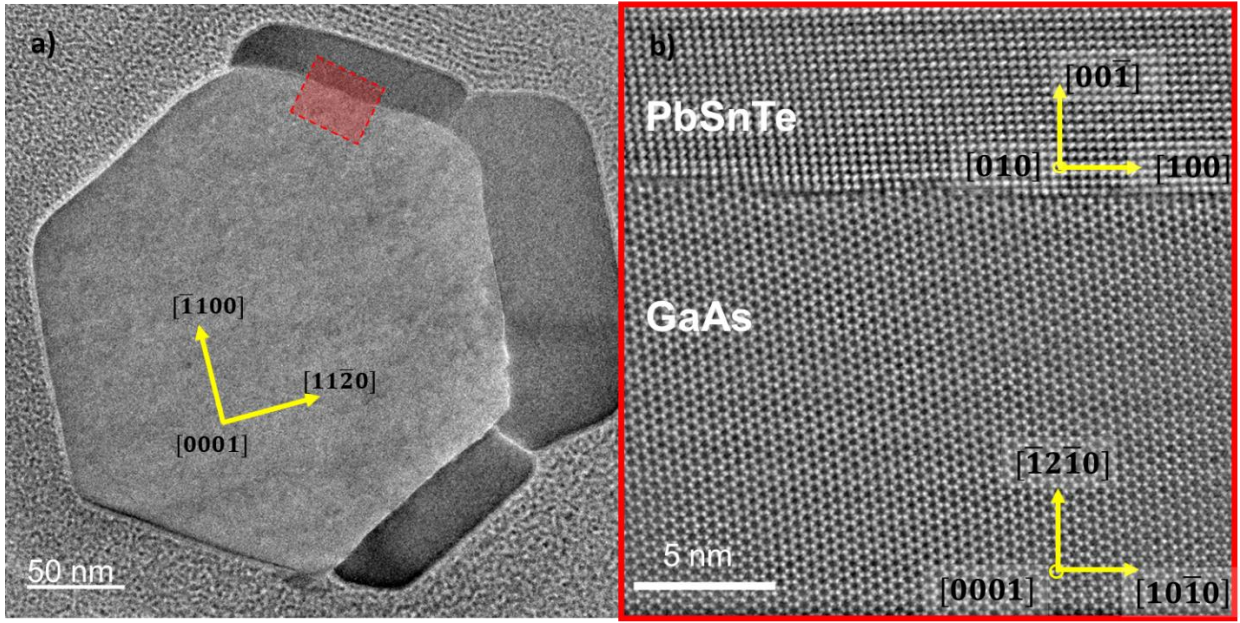

**Figure 10.** TEM images of GaAs(core)-PbSnTe(half-shell) NW cross-section. (a) – low resolution image of the entire cross section; (b) high-resolution image of the core-shell interface taken at the region marked with red rectangle in panel (a).

### 5. Conclusions:

In summary, we have shown that PbTe, SnTe, and Pb$_{1-x}$Sn$_x$Te narrow bandgap semiconductors can be grown as full and half-shells on the sidewalls of WZ GaAs NWs. Contrary to the uniform IV-VI NWs, this kind of hybrid NWs are easier to grow with larger lengths (determined by the III-V core NWs) which is beneficial for further studies of charge carriers transport involving topologically protected surface states occurring at {100} and {110} surfaces of Pb$_{1-x}$Sn$_x$Te above some critical concentration of Sn content. Among the contenders, the binary PbTe shells and $Pb_{0.44}Sn_{0.56}Te$ half shells are best lattice-matched to the {11-20} sidewalls of WZ GaAs NWs, and the lattice mismatch is less than 1.7% and 1.9% respectively, along the [0001] core NW axis. The shells on the inclined NWs are smoother with smattering cracks in comparison to the perpendicular ones. The occurrence of the two types of sidewalls in WZ GaAs NWs enables selection of the orientation of IV-VI NW shells – the {100} IV-VI planes are best fitted to {11-20} WZ GaAs NW sidewalls, whereas the {110} planes are fitted to the {1-100} ones. The latter opens possibilities for investigations of topological {110} surfaces of IV-VI TCI; not investigated yet. To our best knowledge the planar growth of narrow bandgap IV-VI chalcogenides in (110) orientation has not been reported so far, and theoretically predicted topological states on this surface are yet to be experimentally confirmed.

**Corresponding Authors**

*Janusz Sadowski ([janusz.sadowski@ifpan.edu.pl](mailto:janusz.sadowski@ifpan.edu.pl)) *Sania Dad ([saniad@ifpan.edu.pl](mailto:saniad@ifpan.edu.pl))

**Acknowledgment**

The authors acknowledge funding from the National Science Centre Poland, through the projects No: 2019/35/B/ST3/03381, 2019/35/B/ST5/03434, and 2017/27/B/ST3/02470. We thank Prof. T. Story for critical reading of the manuscript-

**Abbreviations**

NWs, nanowires; 1D, one dimensional; MBE, molecular beam epitaxy; TCI, topological crystalline insulator; TI, topological insulator; VLS, vapor liquid solid; HR-TEM, high-resolution transmission electron microscopy; SEM, scanning electron microscope; WZ, wurtzite.

## References:

1.  Hull, R. and J.C. Bean, *Misfit dislocations in lattice-mismatched epitaxial films.* Critical Reviews in Solid State and Material Sciences, 1992. **17**(6): p. 507-546.

2.  Jain, S., A. Harker, and R. Cowley, *Misfit strain and misfit dislocations in lattice mismatched epitaxial layers and other systems.* Philosophical Magazine A, 1997. **75**(6): p. 1461-1515.

3.  Kunert, B., et al., *How to control defect formation in monolithic III/V hetero-epitaxy on (100) Si? A critical review on current approaches.* Semiconductor Science and Technology, 2018. **33**(9): p. 093002.

4.  Chandra, Y., E.S. Flores, and S. Adhikari, *Buckling of 2D nano hetero-structures with moire patterns.* Computational Materials Science, 2020. **177**: p. 109507.

5.  Sugumaran, P.J., J. Zhang, and Y. Zhang, *Synthesis of stable core-shell perovskite based nano-heterostructures.* Journal of Colloid and Interface Science, 2022. **628**: p. 121-130.

6.  Hasan, M.Z. and C.L. Kane, *Colloquium: topological insulators.* Reviews of modern physics, 2010. **82**(4): p. 3045.

7.  Qi, X.L. and S.C. Zhang, *Topological insulators and superconductors.* Reviews of Modern Physics, 2011. **83**(4).

8.  Springholz, G., *Molecular Beam Epitaxy of IV–VI Semiconductors: Fundamentals, Low-dimensional Structures, and Device Applications*, in *Molecular Beam Epitaxy*. 2018, Elsevier. p. 211-276.

9.  Liu, H., et al., *Photothermoelectric SnTe photodetector with broad spectral response and high on/off ratio.* ACS Applied Materials & Interfaces, 2020. **12**(44): p. 49830-49839.

10. Safaei, S., P. Kacman, and R. Buczko, *Topological crystalline insulator (Pb, Sn) Te: Surface states and their spin polarization.* Physical Review B, 2013. **88**(4): p. 045305.

11. Fu, L., *Topological crystalline insulators.* Physical Review Letters, 2011. **106**(10): p. 106802.

12. Hsieh, T.H., et al., *Topological crystalline insulators in the SnTe material class.* Nature communications, 2012. **3**(1): p. 1-7.

13. Tanaka, Y., et al., *Experimental realization of a topological crystalline insulator in SnTe.* Nature Physics, 2012. **8**(11): p. 800-803.

14. Xu, S.-Y., et al., *Observation of a topological crystalline insulator phase and topological phase transition in Pb1− xSnxTe.* Nature communications, 2012. **3**(1): p. 1-11.

15. Dziawa, P., et al., *Topological crystalline insulator states in Pb 1− x Sn x Se.* Nature materials, 2012. **11**(12): p. 1023-1027.

16. Krizman, G., et al., *Tunable Dirac interface states in topological superlattices.* Physical Review B, 2018. **98**(7): p. 075303.

17. Assaf, B., et al., *Massive and massless Dirac fermions in Pb 1− x Sn x Te topological crystalline insulator probed by magneto-optical absorption.* Scientific reports, 2016. **6**(1): p. 1-8.

18. Schindler, F., et al., *Higher-order topological insulators.* Science advances, 2018. **4**(6): p. eaat0346.

19. Schindler, F., et al., *Higher-order topology in bismuth.* Nature physics, 2018. **14**(9): p. 918-924.

20. Bukała, M., et al., *Stability of III–V and IV–VI nanowires—A theoretical study.* Physica E: Low-dimensional Systems and Nanostructures, 2010. **42**(4): p. 795-798.

21.    Wu, Y., et al., *Controlled growth and structures of molecular-scale silicon nanowires.* Nano letters, 2004. **4**(3): p. 433-436.

22.    Hersee, S.D., X. Sun, and X. Wang, *The controlled growth of GaN nanowires.* Nano letters, 2006. **6**(8): p. 1808-1811.

23.    Bauer, B., et al., *Position controlled self-catalyzed growth of GaAs nanowires by molecular beam epitaxy.* Nanotechnology, 2010. **21**(43): p. 435601.

24.    Yang, P., et al., *Controlled growth of ZnO nanowires and their optical properties.* Advanced functional materials, 2002. **12**(5): p. 323-331.

25.    Wang, F., A. Dong, and W.E. Buhro, *Solution–liquid–solid synthesis, properties, and applications of one-dimensional colloidal semiconductor nanorods and nanowires.* Chemical reviews, 2016. **116**(18): p. 10888-10933.

26.    Wagner, a.R. and s.W. Ellis, *Vapor-liquid-solid mechanism of single crystal growth.* Applied physics letters, 1964. **4**(5): p. 89-90.

27.    Dziawa, P., et al., *Defect free PbTe nanowires grown by molecular beam epitaxy on GaAs (111) B substrates.* Crystal growth & design, 2010. **10**(1): p. 109-113.

28.    Panciera, F., et al., *Phase selection in self-catalyzed GaAs nanowires.* Nano letters, 2020. **20**(3): p. 1669-1675.

29.    Parameshwaran, V. and P. Taylor, *Alloying Behavior and Crystallinity of (111)-Oriented Lead Tin Telluride Grown on (100)-Oriented Gallium Arsenide.* 2019, CCDC Army Research Laboratory.

30.    Yang, L., et al., *Novel route to scalable synthesis of II–VI semiconductor nanowires: Catalyst-assisted vacuum thermal evaporation.* Journal of crystal growth, 2010. **312**(20): p. 2852-2856.

31.    Dick, K.A., *A review of nanowire growth promoted by alloys and non-alloying elements with emphasis on Au-assisted III–V nanowires.* Progress in Crystal Growth and Characterization of Materials, 2008. **54**(3-4): p. 138-173.

32.    Wang, J., et al., *Core/shell colloidal quantum dot exciplex states for the development of highly efficient quantum-dot-sensitized solar cells.* Journal of the American Chemical Society, 2013. **135**(42): p. 15913-15922.

33.    Reiss, P., M. Protiere, and L. Li, *Core/shell semiconductor nanocrystals.* small, 2009. **5**(2): p. 154-168.

34.    Gao, P.X., et al., *Metal/semiconductor core/shell nanodisks and nanotubes.* Advanced Functional Materials, 2006. **16**(1): p. 53-62.

35.    Brumer, M., et al., *PbSe/PbS and PbSe/PbSexS1–x core/shell nanocrystals.* Advanced Functional Materials, 2005. **15**(7): p. 1111-1116.

36.    Kockert, M., et al., *Semimetal to semiconductor transition in Bi/TiO 2 core/shell nanowires.* Nanoscale Advances, 2021. **3**(1): p. 263-271.

37.    Tang, X., et al., *Single halide perovskite/semiconductor core/shell quantum dots with ultrastability and nonblinking properties.* Advanced Science, 2019. **6**(18): p. 1900412.

38.    Sköld, N., et al., *Growth and Optical Properties of Strained GaAs− Ga x In1-x P Core− Shell Nanowires.* Nano letters, 2005. **5**(10): p. 1943-1947.

39.    Cirloganu, C.M., et al., *Enhanced carrier multiplication in engineered quasi-type-II quantum dots.* Nature communications, 2014. **5**(1): p. 1-8.

40.    Miranti, R., et al., *Exclusive electron transport in Core@ Shell PbTe@ PbS colloidal semiconductor nanocrystal assemblies.* ACS nano, 2020. **14**(3): p. 3242-3250.

41.    Aryal, S. and R. Pati, *PbTe (core)/PbS (shell) Nanowire: Electronic Structure, Thermodynamic Stability, and Mechanical and Optical Properties.* The Journal of Physical Chemistry C, 2021.

42.    Koenigsmann, C., et al., *Designing enhanced one-dimensional electrocatalysts for the oxygen reduction reaction: Probing size-and composition-dependent electrocatalytic behavior in noble metal nanowires.* The Journal of Physical Chemistry Letters, 2012. **3**(22): p. 3385-3398.

43.    Percival, S.J. and B. Zhang, *Electrocatalytic reduction of oxygen at single platinum nanowires.* The Journal of Physical Chemistry C, 2013. **117**(27): p. 13928-13935.

44.    Zhang, H., B. Man, and Q. Zhang, *Topological crystalline insulator SnTe/Si vertical heterostructure photodetectors for high-performance near-infrared detection.* ACS Applied Materials & Interfaces, 2017. **9**(16): p. 14067-14077.

45.    Ginting, D., et al., *Enhancement of thermoelectric performance in Na-doped Pb0. 6Sn0. 4Te0. 95–x Se x S0. 05 via breaking the inversion symmetry, band convergence, and nanostructuring by multiple elements doping.* ACS applied materials & interfaces, 2018. **10**(14): p. 11613-11622.

46.    Haidet, B.B., et al., *Interface structure and luminescence properties of epitaxial PbSe films on InAs (111) A.* Journal of Vacuum Science & Technology A: Vacuum, Surfaces, and Films, 2021. **39**(2): p. 023404.

47.    Sadowski, J. and M. Herman, *Hard heteroepitaxy of molecular beam epitaxial grown PbTe on off oriented GaAs (100) substrates.* Journal of crystal growth, 1995. **146**(1-4): p. 449-454.

48.    Liu, X., et al., *Unraveling the structural and electronic properties of strained PbSe on GaAs.* Journal of Crystal Growth, 2021. **570**: p. 126235.

49.    Yan, X., et al., *Analysis of critical dimensions for nanowire core-multishell heterostructures.* Nanoscale Research Letters, 2015. **10**(1): p. 1-7.

50.    Sadowski, J., et al., *Defect-free SnTe topological crystalline insulator nanowires grown by molecular beam epitaxy on graphene.* Nanoscale, 2018. **10**(44): p. 20772-20778.

51.    Nguyen, N.M., W. Brzezicki, and T. Hyart, *Corner states, hinge states, and Majorana modes in SnTe nanowires.* Physical Review B, 2022. **105**(7): p. 075310.

52.    Jacobsson, D., et al., *Phase transformation in radially merged wurtzite GaAs nanowires.* Crystal growth & design, 2015. **15**(10): p. 4795-4803.

53.    Bauer Pereira, P., et al., *Lattice dynamics and structure of GeTe, SnTe and PbTe.* physica status solidi (b), 2013. **250**(7): p. 1300-1307.

54.    Volobuev, V.V., et al., *Giant Rashba splitting in Pb1–xSnxTe (111) topological crystalline insulator films controlled by Bi doping in the bulk.* Advanced Materials, 2017. **29**(3): p. 1604185.

55.    Wagner, J.W. and J.C. Woolley, *Phase studies of the Pb1− xSnxTe alloys.* Materials Research Bulletin, 1967. **2**(11): p. 1055-1062.